\documentclass{article}

\usepackage{microtype}
\usepackage{graphicx}
\usepackage{subcaption}
\usepackage{booktabs} % for professional tables
\usepackage{multirow} % for multirow cells in tables
\usepackage{tabularx}
\usepackage{hyperref}
\usepackage{array}
\usepackage{listings}
\usepackage{float}

\usepackage{tcolorbox}
\usepackage{fancyvrb}
\tcbset{colback=gray!5,colframe=gray!50,boxrule=0.5pt,arc=2pt}

\usepackage[preprint]{icml2026}

\usepackage{amsmath}
\usepackage{amssymb}
\usepackage{mathtools}
\usepackage{amsthm}

\usepackage[capitalize,noabbrev]{cleveref}

\theoremstyle{plain}

\theoremstyle{definition}

\theoremstyle{remark}

\usepackage[textsize=tiny]{todonotes}

\begin{document}

\twocolumn[
  \icmltitle{BRITE: A Benchmark for Reliable and Interpretable T2V Evaluation on Implausible Scenarios}

  %\icmltitle{Human Centered Framework for T2V Evaluation using Off-Manifold Prompts}

 % List of affiliations: The first argument should be a (short) identifier you
  % will use later to specify author affiliations.
  % Affiliations will be numbered in order of appearance.

  \begin{icmlauthorlist}
    \icmlauthor{Advait Tilak}{isu}
    \icmlauthor{Jiwon Choi}{isu}
    \icmlauthor{Nazifa Mouli}{isu}
    \icmlauthor{Wei Le}{isu}
    
  \end{icmlauthorlist}

  \icmlaffiliation{isu}{Department of Computer Science, Iowa State University, Ames, Iowa, USA}

  \icmlcorrespondingauthor{Advait Tilak}{advait46@iastate.edu}

  % You may provide any keywords that you find helpful for describing your
  % paper; these are used to populate the "keywords" metadata in the PDF.
  \icmlkeywords{Machine Learning, Generative AI, Text-to-Video, Benchmark, Vision-Language Models, Evaluation}

  \vskip 0.3in
]

% This must go after the closing bracket ] following \twocolumn[ ...

% This command actually creates the footnote in the first column listing the
% affiliations and the copyright notice. The command takes one argument, which
% is text to display at the start of the footnote. 
% If you have no special notice, KEEP empty braces:
\printAffiliationsAndNotice{}  % no special notice (required even if empty)

\begin{abstract}

The rapid advancement of photorealistic Text-to-Video (T2V) generation brings in an urgent need for up-to-date evaluation methods. Existing benchmarks largely overlooked implausible scenarios and do not measure audio-visual alignment. We introduce BRITE, the first framework that unifies (1) implausible prompting,  (2) fine-grained assessment of audio-visual consistency, and (3) QA-based interpretable evaluation into a comprehensive T2V benchmark. Unlike fully automated Multimodal LLM-based pipelines, which are prone to hallucination and prompt ambiguity, BRITE guarantees reliability through a rigorous human-in-the-loop protocol for benchmark creation.  Evaluating five state-of-the-art models (Sora 2, Veo 3.1, Runway Gen4.5, Pixverse V5.5, and Qwen3Max), we reveal a critical performance gap: while models excel at static object composition, they exhibit significant degradation in object-action binding and audio-visual synchronization. Our framework offers the community a reliable, interpretable benchmark and evaluation framework that can detect and locate limitations in the next generation of T2V models, especially for off-manifold prompts. We release our code, prompts and data at 
\href{https://doi.org/10.6084/m9.figshare.31179547}{https://doi.org/10.6084/m9.figshare.31179547}.

\end{abstract}

\section{Introduction}
Recent advancements in video generation have achieved remarkable visual realism and audio-visual synchronization. However, the rapid scaling of Text-to-Video (T2V) capabilities has far outpaced evaluation methodologies, creating a critical bottleneck in measuring true model intelligence. While early benchmarks focused on low-level visual quality, such as sharpness and temporal smoothness, recent research now emphasizes complex scene structure, action binding, and object interaction \cite{sun2025t2vcompbench, huang2024vbench}. Despite these improvements, most existing benchmarks \cite{bansal2025videophy2, chen2025t2vworldbench} evaluated realism under the "forward-generation" question: "Given premise $X$, can the model produce a plausible and coherent video?" This focus leaves two critical gaps: first, high realism scores often mask poor instruction faithfulness, as models may reinforce learned real-world priors; second, these benchmarks typically treat videos as silent sequences, failing to measure audio-visual synchronization.

This motivates a shift toward a "reverse" evaluation perspective: assessing model behavior when the premise itself is unrealistic. Evaluating these implausible scenarios is critical because T2V models, largely trained on internet-scale data, tend to regress toward realistic manifolds when faced with implausible prompts \cite{bai2025impossible}. In such cases, the generative process collapses toward learned real-world priors rather than constructing a coherent world that adheres to rule-violating prompt instructions. While recent work has begun exploring such implausible prompts, existing evaluation~\cite{bai2025impossible} suffers from systemic limitations that hinder true model diagnosis. Current evaluation \cite{bai2025impossible} relies on holistic judgments or a single "pass-fail" score----merely checking if an impossible event "happens," these metrics prevent researchers from pinpointing specifically where the generative process fails. Furthermore, the reliance on automated LLM judges can introduce a circular evaluation bias that undermines the assessment of off-manifold content.

\begin{table*}[t]
\caption{Comparison of existing T2V benchmarks and our method across prompt design, evaluation targets, methodologies, and diagnostic capabilities.}
\centering
\small
\setlength{\tabcolsep}{4pt}
\renewcommand{\arraystretch}{1.1}

\begin{tabular}{c | c | c c c | c c c | c c}
\hline
\\[-0.99em]  % add breathing space below top border

\multirow{2}{*}{Benchmark} 
& \multirow{2}{*}{\shortstack{Implausible\\scenario}}
& \multicolumn{3}{c|}{Evaluation Target} 
& \multicolumn{3}{c|}{Evaluation Methodology} 
& \multicolumn{2}{c}{Diagnostics} \\

\cline{3-5}
\cline{6-8}
\cline{9-10}

& 
& Visual & Audio & \shortstack{AV\\Sync}
& \shortstack{Atomic\\QA} 
& \shortstack{Human\\Q review} 
& \shortstack{Human\\Q answer}
& \shortstack{Failure\\localization} 
& \shortstack{Yes/No\\accuracy} \\
\hline

VBench++\cite{huang2024vbenchplus}
& $\times$ 
& \checkmark & $\times$ & $\times$ 
& $\times$ & $\times$ & $\times$ 
& \checkmark & $\times$ \\

EvalCrafter\cite{liu2024evalcrafter}
& $\times$ 
& \checkmark & $\times$ & $\times$ 
& $\times$ & \checkmark & $\times$ 
& $\times$ & $\times$ \\

T2V-CompBench\cite{sun2025t2vcompbench} 
& $\times$ 
& $\times$ & $\times$ & $\times$  
& $\times$ & \checkmark & $\times$ 
& \checkmark & $\times$ \\

ETVA\cite{guan2025etva}
& $\times$ 
& $\times$ & $\times$ & $\times$ 
& \checkmark & $\times$ & $\times$ 
& \checkmark & \checkmark \\

VideoPhy \cite{bansal2025videophy2}
& $\times$ 
& $\times$ & $\times$ & $\times$ 
& $\times$ & \checkmark & \checkmark 
& $\times$ & $\times$ \\

T2VWorldBench \cite{chen2025t2vworldbench}
& $\times$ 
& \checkmark & $\times$ & $\times$ 
& $\times$ & \checkmark & $\times$ 
& $\times$ & $\times$ \\

IPV-Bench \cite{bai2025impossible}
& \checkmark 
& \checkmark & $\times$ & $\times$ 
& $\times$ & \checkmark & $\times$
& $\times$ & $\times$ \\

\hline
\textbf{Our work} 
& \checkmark 
& \checkmark & \checkmark & \checkmark 
& \checkmark & \checkmark & \checkmark 
& \checkmark & \checkmark \\
\hline
\end{tabular}
\label{tab:novelty_analysis}
\end{table*}

To bridge this gap, we propose BRITE, a benchmark and human-centric evaluation framework that systematically assess models' performance on implausible scenarios. In addition to add more implausible categories on top of prior work~\cite{bai2025impossible}, our benchmark is also the first that introduced the audio and audio-visual synchronization assessment for T2V models. Importantly, we developed fine-grained QA based evaluation, following~\cite{guan2025etva} and generated fine-grained questions based on a diverse of dimensions, which can help locate the root causes of failures. Our video generation and question generation used LLMs with human-in-the-loop to reduce the noise in the benchmark. We used human annotators to answer questions instead of LLM judges to avoid potential circular biases mentioned above.

Our contributions are as follows:
\begin{itemize}

    \item We developed a comprehensive T2V evaluation framework for implausible scenarios: the benchmark is {\it reliable} and we used human-in-the-loop to remove noisy data in different stages of benchmark creation; it is {\it interpretable} in that we developed fine-grained questions to help localize where and how a model fails.
    
    \item To the best of our knowledge, it is the first T2V benchmark that evaluates the correctness of the audio and audio-visual synchronization, measuring whether generated sounds are temporally aligned with specific visual actions (e.g., footsteps matching the pace of walking). See Table~\ref{fig:T2V_framework} for further details of our novelty.
    
    \item We evaluated five SOTA models, including Runway Gen 4.5, Sora 2, and Veo 3, across 500 videos and 1,364 questions. Our results showed that there exist performance gaps: models succeed at static tasks like subject generation, but consistently fail at dynamic tasks like action generation. Furthermore, while they generate semantically correct audio, they fail to synchronize that sound with the video's visual timing.

    %\wei{a clear divide in capability}

  %  Sora 2, Veo 3.1, Runway Gen4.5, PixVerse v5.5, and Qwen3Max
    
   \end{itemize}

\section{An Overview}

\begin{figure*}[t]
    \centering
    \includegraphics[width=0.98\linewidth]{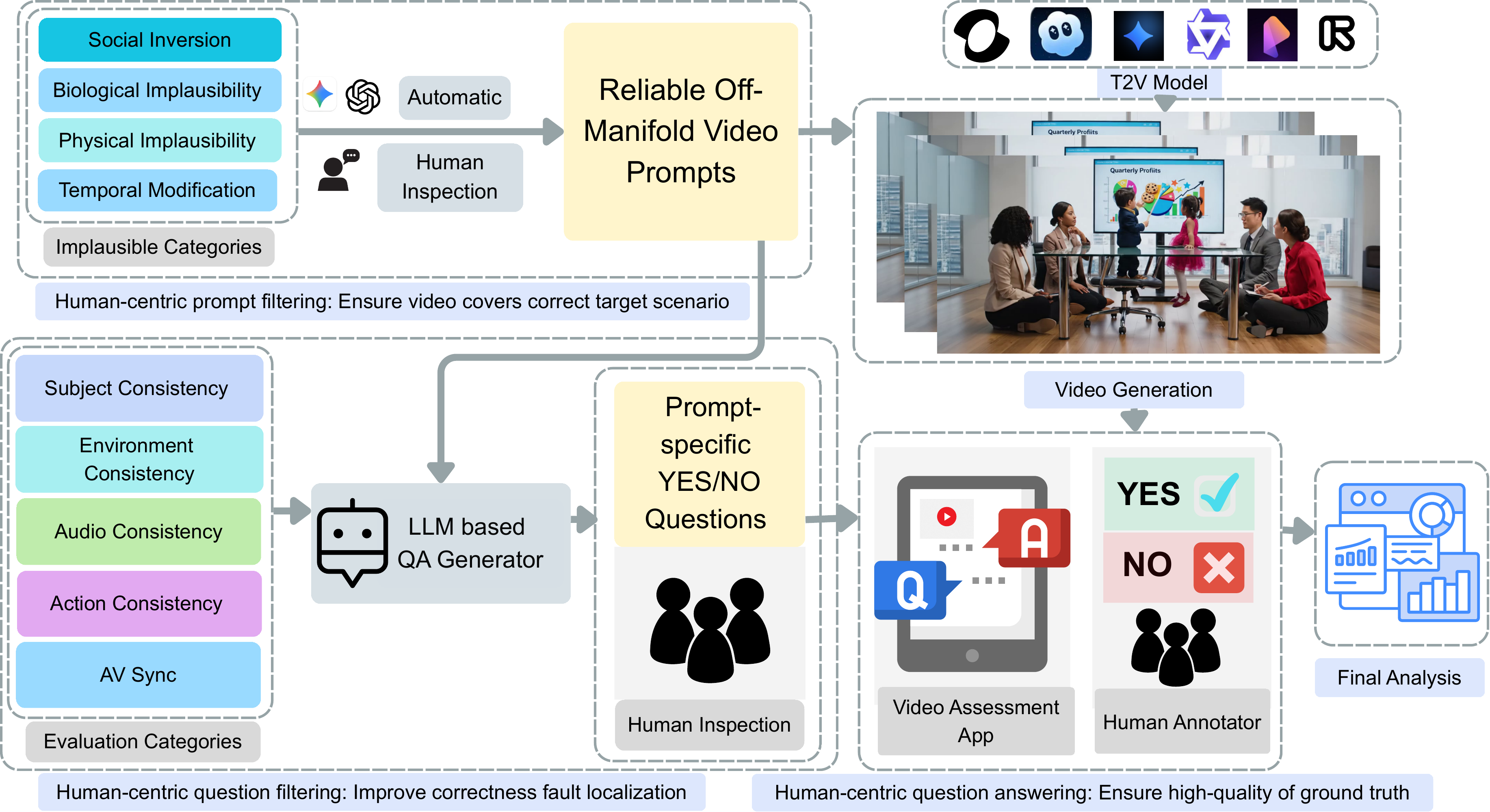}
    \caption{A Reliable and Interpretable Benchmark Generation for T2V Evaluation: An Overview Framework}
    \label{fig:T2V_framework}
\end{figure*}

%\wei{Overview figure: Nazifa - TODO}
%We introduce BRITE Bench, a comprehensive framework designed to evaluate Text-to-Video (T2V) model adherence to implausible prompts. Unlike standard benchmarks that prioritize plausibility, our framework provides an interpretable, fine-grained diagnosis of instruction following under off-manifold conditions. 

Figure~\ref{fig:T2V_framework} presents an overview of our work. The evaluation pipeline begins with \textbf{Prompt Curation}, where deliberate rule violations are categorized into four Implausible Categories: Social Inversion, Biological Implausibility, Physical Implausibility, and Temporal Modification. As detailed in \cref{prompts}, these categories serve as the foundation for testing a model’s ability to prioritize explicit instructions over learned world priors. We further refine these inputs through human-centric prompt filtering (\cref{prompt_filtering}) to ensure that each prompt accurately represents the target off-manifold scenario and can be evaluated objectively.

Following curation, these prompts are used for \textbf{Video Generation} to produce a video $v$ from a T2V model. To facilitate a structured evaluation, an LLM-based QA Generator derives a set of prompt-specific binary questions $\mathcal{Q} = \{Q_1, Q_2, \dots, Q_n\}$ from the original prompt (\cref{questions}). These questions correspond to single, verifiable claims regarding prompt adherence and undergo human-centric filtering (\cref{question_filtering}) to improve fault localization. During this stage, questions are assigned to specific Evaluation Categories: Subject, Action, Environment, Audio, and Audio-Visual Synchronization.

Subsequently, the \textbf{Video Assessment Protocol} (\cref{human_assesment_protocol}) resolves each question through human inspection using a structured assessment application. Judgments are based on integrated sources: the generated video $v$ and the specific atomic question $Q_i$. In the  \textbf{Final Analysis} (\cref{final_analysis}), these binary judgments are aggregated to compute dimension-wise adherence scores and overall benchmark metrics. This formulation is specifically designed to measure instruction dominance versus prior dominance by evaluating whether the generated video realizes requested rule violations while maintaining its internal coherence.

\section{Benchmark Construction}

\subsection{Implausible Prompt Generation and Categories}
\label{prompts}
\begin{figure*}[t]
    \centering
    \includegraphics[width=\linewidth]{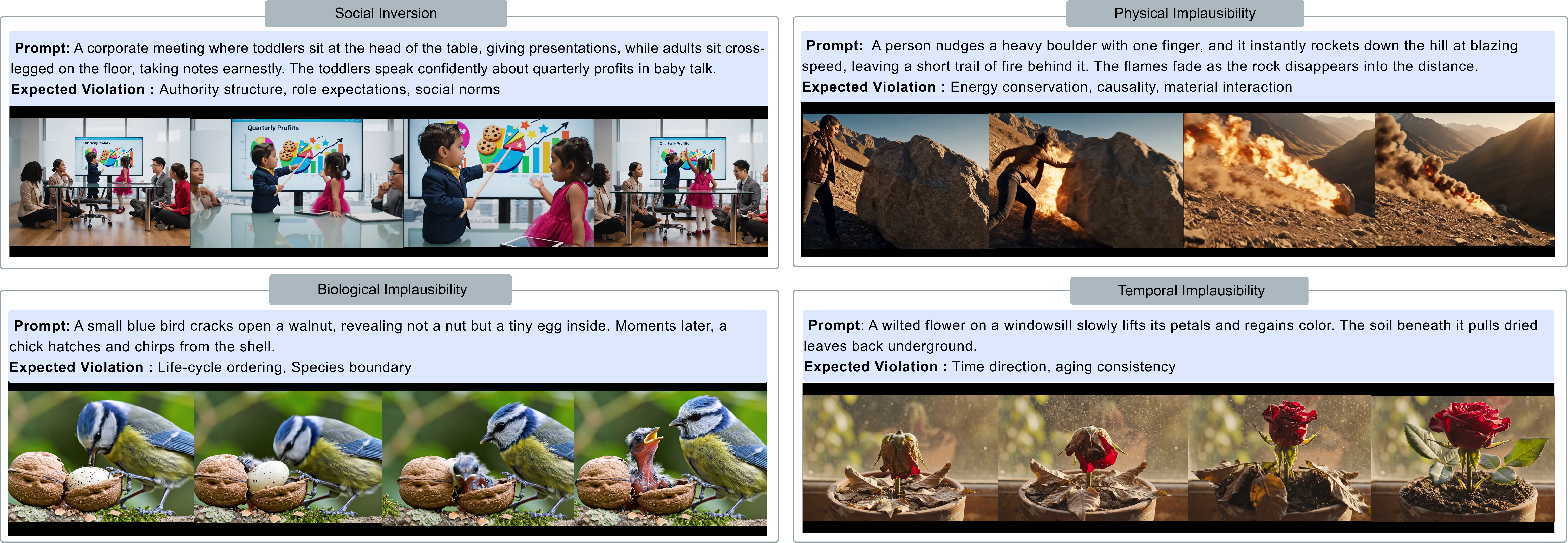}
    \caption{Examples from BRITE across four implausibility categories. Each example pairs an implausible prompt with generated video frames (we selected {\it correct} examples) and the expected violated world rules, covering social inversion, biological implausibility, physical implausibility, and temporal implausibility.}
    \label{fig:FantasticBench_example}
\end{figure*}

Table \ref{tab:implausibility-taxonomy} presents implausible scenarios we aim to evaluate, and Figure~\ref{fig:FantasticBench_example} provides an example for each scenario. Specifically, 
{\it Social Inversion}  evaluates scenes reversing societal roles or hierarchies (e.g., a patient diagnosing a doctor) while maintaining physical realism to test the model's ability to decouple social norms from visual priors.
{\it Biological Implausibility} evaluates organisms acting against fundamental anatomy or habitat constraints (e.g., a barking cat or roots growing skyward) to evaluate the override of species-specific behaviors. {\it Physical Implausibility} evaluates Intentional violations of physical laws such as gravity or material properties while maintaining entity realism (e.g., a person floating upward upon jumping from a cliff). {\it Temporal Modification} alterates the linear flow of time (e.g., shattered glass reassembling). 

We prompted GPT-4 and Gemini 2.5 Pro to synthesize prompts for video generation. Our prompt to GPT and Gemini contained the following: the description of "implausible scenarios" (see {\it violated rules} in Table~\ref{fig:FantasticBench_example}), a manually created short example video generation prompt, e.g., "a cat barking ...", and the number of video generation prompts we aim to generate for each category and their lengths. 
See Appendix~\ref{app:meta_prompt} for an example.The generated prompts were then filtered by human annotators before sending for video generation (Section \cref{prompt_filtering}).

%Unlike IPV-Bench, which subordinates temporal anomalies under general physical conservation laws, we isolate Temporal Modification as a distinct primary category. This allows us to specifically evaluate whether models can decouple causal reasoning from their training bias toward linear temporal progression.

%\wei{TODO: need to add examples for Table 1 Nazifa}
\begin{table*}[t]
\caption{Implausibility categories, violated rules, and the corresponding model limitations explored by our benchmark.}
\centering
\small
\setlength{\tabcolsep}{6pt}
\renewcommand{\arraystretch}{1.15}
\begin{tabular}{
    >{\centering\arraybackslash}m{2.2cm}
    m{6cm}
    >{\raggedright\arraybackslash}m{6cm}
}
\toprule
\textbf{Category} & \textbf{Violated Rules} & \textbf{Explored Limitation} \\
\midrule

\textbf{Social Inversion}
& Authority Structure, Role Expectations, Social Norms, Service Hierarchies
& Ability to invert social roles and norms without altering physical or biological consistency. \\

\midrule

\textbf{Biological Implausibility}
& Species-specific Behavior, Species Boundaries, Physiology, Habitat Constraints, Life-cycle Ordering
& Ability to override biological constraints independently of visual appearance and physical realism. \\

\midrule

\textbf{Physical Implausibility}
& Gravity, Material Interaction, Causality, Collision, Thermal Dynamics, Energy Conservation
& Ability to violate physical laws while preserving entity identity and overall scene coherence. \\

\midrule

\textbf{Temporal Modification}
& Time Direction, Rate of Change, Temporal Synchrony, Aging Consistency
& Ability to maintain temporal coherence under altered time-flow assumptions. \\
\bottomrule
\end{tabular}
\label{tab:implausibility-taxonomy}
\end{table*}

\subsection{Atomic Question Generation} \label{questions}
For each video generation prompt, we produce a collection of atomic "Yes/No" questions targeting distinct aspects of prompt instruction adherence. To ensure a fine-grained evaluation, we avoid high-level general question like "Does this violate gravity?") but favor questions that isolate specific attributes, transitions, and interactions. For example, given a video generation prompt regarding a reversing flame, we decompose the evaluation into specific, atomic questions: “Does the ash texture revert to wood?” “Does the flame travel toward the match head?”, and “Does the flame diminish as it reaches the tip?”

\begin{figure}[t]
    \centering
    \includegraphics[width= \columnwidth]{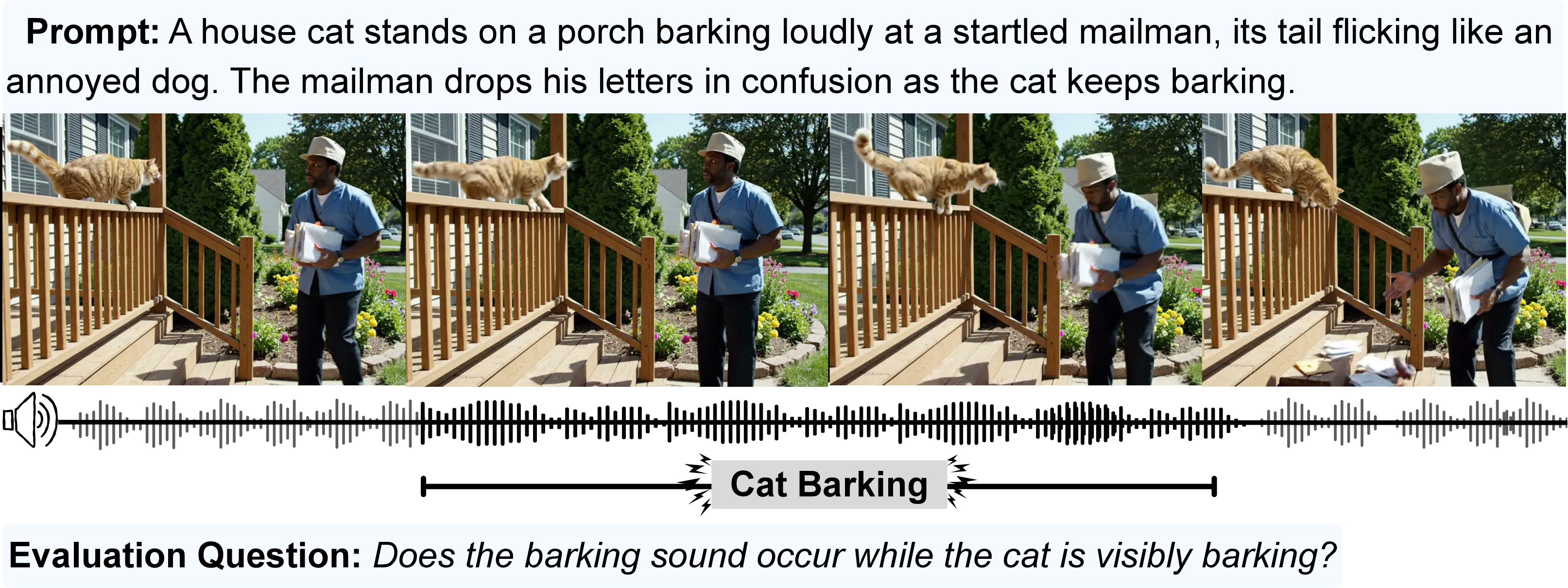}
    \caption{Example of atomic question generation for Audio-visual consistency (Cat Barking Scenario)}
    \label{fig:atomic_example}
\end{figure}

These questions are automatically generated by Gemini 2.5 Pro. In the prompt to Gemini,  we guided the models using one of the following five dimensions: Subject, Action, Environment, Audio, and {Audio-Visual Syncronization}. This decomposition ensures that a model is not penalized for its visual quality and we are evaluating different aspects of video semantics. To the best of our knowledge, BRITE is the first that extends evaluation beyond the visual modality. By introducing explicit Audio and Audio-Visual Synchronization dimensions, we address the limitations in current baselines~\cite{liu2024evalcrafter,sun2025t2vcompbench,guan2025etva, bai2025impossible,bansal2025videophy2,huang2024vbench,chen2025t2vworldbench}, quantifying not just visual hallucinations, but also the {correctness of the sound} and timing of generated sound. Figure~\ref{fig:atomic_example} provides such an example.

In our prompt to Gemini 2.5, we first given detailed information about each of the 4 implausible categories and 5 dimensions. We then prompted the LLM to generate 3-5 questions per evaluation dimension per prompt and output them in a structured JSON format (see Appendix~\ref{app:question_generation} for our question generation prompt). This process was carried out on a (video generation) prompt by prompt basis to avoid overloading the context window and prevent hallucinations. These questions were then filtered by human annotators (\cref{question_filtering}).

%In total, BRITE comprises 1,364 atomic questions, creating a rigorous evaluation environment that necessitates grounded causal understanding over superficial pattern matching.

\section{A Human-Centric Evaluation Framework}
\iffalse

Human-centric: more reliable and 

Also : Our evaluation in based on implausible scenarios. I fell MLLM based evaluators will find it very difficult to evaluate such scenarios. I am not sure if they have the ability to differentiate between actual hallucinations and correct alignment with our prompts.

\fi

%\wei{jiwon}
\iffalse
We propose a human-centric framework for evaluating instruction following in Text-to-Video (T2V) models under off-manifold conditions. Prompt adherence is assessed via atomic, interpretable Yes/No judgments rather than a single holistic score. This enables reliable and fine-grained diagnosis while avoiding circular model-based evaluation.
\fi
%Our framework incorporates human verification at three critical stages: prompt filtering, question filtering, and answer validation. Each stage addresses a distinct failure mode of automated or model-based evaluation pipelines, particularly under off-manifold conditions.

\subsection{Human-Centric Prompt Filtering} \label{prompt_filtering}
%\wei{what types of mistakes can exist in prompts: add note past LLM based approach generate images instead of videos }

%Prompts are automatically generated by LLMs. There are potential problems on these Prompts.  

\begin{table}[H]
\centering
\caption{Examples of Filtered Prompts}
\label{tab:prompt_filtering_examples}
\footnotesize
\setlength{\tabcolsep}{4pt}
\renewcommand{\arraystretch}{1.15}
\begin{tabular}{p{0.26\columnwidth} p{0.70\columnwidth}}
\toprule
\textbf{Failure Type} & \textbf{Prompt Example} \\
\midrule
% Old ambiguity example:
%\textbf{Ambiguity} &
%A dog is told to \textit{throw} a ball. \\
%\textbf{Ambiguity} &
%The flower’s sadness seems \textit{realistic}. \\

\textbf{Ambiguity} &
A dog speaks English with a \textit{natural} voice. \\
\textbf{Unverifiable} &
Honey takes \textit{hours} to fall onto toast. \\
\textbf{Confounded} &
A cat \textit{barks} while the scene \textit{remains frozen}. \\
\bottomrule
\end{tabular}
\end{table}

%\wei{better examples}

%We performed manual evaluation and filtered the three types of 
%In the prior work, without prompt filtering, we generate
Our human-in-the-loop protocol first involves an inspection of video generation prompts. Specifically, we filtered three types of mistakes in a prompt, shown in Table \ref{tab:prompt_filtering_examples}. The first type is {\it semantic ambiguity}. LLMs often generate prompts that are grammatically correct but contain ground truth that is open to subjective interpretation. The second type is {\it unverifiable requirements}----prompts containing constraints that either surpass the capabilities of current T2V models (e.g., video length) or cannot be validated through visual or auditory evidence. Third, it is confounded or extremely implausible. LLMs occasionally stack multiple conflicting rules into a single prompt.

\subsection{Human-Centric Question Filtering} \label{question_filtering} 

Automated question synthesis can also introduce noise that compromises evaluation reliability. Specifically, we filtered out two types of undesired LLM-generated questions. First, LLMs frequently ask questions that are irrelevant or not constrained by the prompt. For instance, given a prompt of "a person jumping into a pool", the LLM may specify questions like "Is the pool in a backyard?". The location of the pool is neither explicitly stated nor logically implied in the prompt. Second, the question does not align with the prompt and can't yield a clearly "yes" answer. For example, an LLM might generate a question "Did the cat quack?" or "Did the cat moo?" for a prompt "the cat is barking". To automatically check the question/answer with the prompt in such cases require 
a sound natural language analysis.  In our design, 
we chose questions that are directly aligned with the prompt, e.g., "did the cat bark?"

%where a "Yes" response would paradoxically indicate a model failure, complicating downstream metric calculation. 

%Second, LLMs often fail to maintain a consistent "Yes-for-Success" framework. For example, an LLM might generate a negative framing (e.g., "Does the cat fail to bark?"), where a "Yes" response would paradoxically indicate a model failure, complicating downstream metric calculation. 

%Similar to prompt generation, automated question synthesis introduces noise that compromises evaluation reliability. We identified two critical failure modes in LLM-generated questions that necessitate human intervention:

%Inconsistent Polarity : LLMs often fail to maintain a consistent "Yes-for-Success" framework. For example, an LLM might generate a negative framing (e.g., "Does the cat fail to bark?"), where a "Yes" response would paradoxically indicate a model failure, complicating downstream metric calculation.

%Over-Specified Constraints : 

%LLMs frequently ask questions that are irrelevant. For instance, given a generic prompt of "a person jumping into a pool", the LLM may specify questions like "Is the pool in a backyard?". The location of the pool is neither explicitly stated nor logically implied in the prompt.

To eliminate these noise and ensure that every question provides a clear, independent failure signal, we employed a Dual-Criterion Filtering Protocol. Human annotators reviewed all candidate questions, retaining only those that satisfied one of two strict logical categories: 

\begin{itemize}
	\item \textbf{Explicit Constraints:} These questions target attributes directly specified in the text prompt.
    
Criterion: Does the question verify a fact that exists textually in the prompt?

	\item \textbf{Implicit Contextual Entailment:} These questions target environmental or physical details that are not explicitly stated but are logically necessary consequences of the prompt components ("World Knowledge").

Criterion: Does the presence of the Subject or Action necessitate this environment?
\end{itemize}
For example, in the prompt "A gazelle chases after a lion", the text does not explicitly describe the environment. However, the presence of these entities logically entails a Savanna-like environment. Therefore, a question assessing whether the background is appropriate is valid, as a mismatch would constitute a failure of contextual reasoning.

\begin{figure}[t]
    \centering    \includegraphics[width=\linewidth]{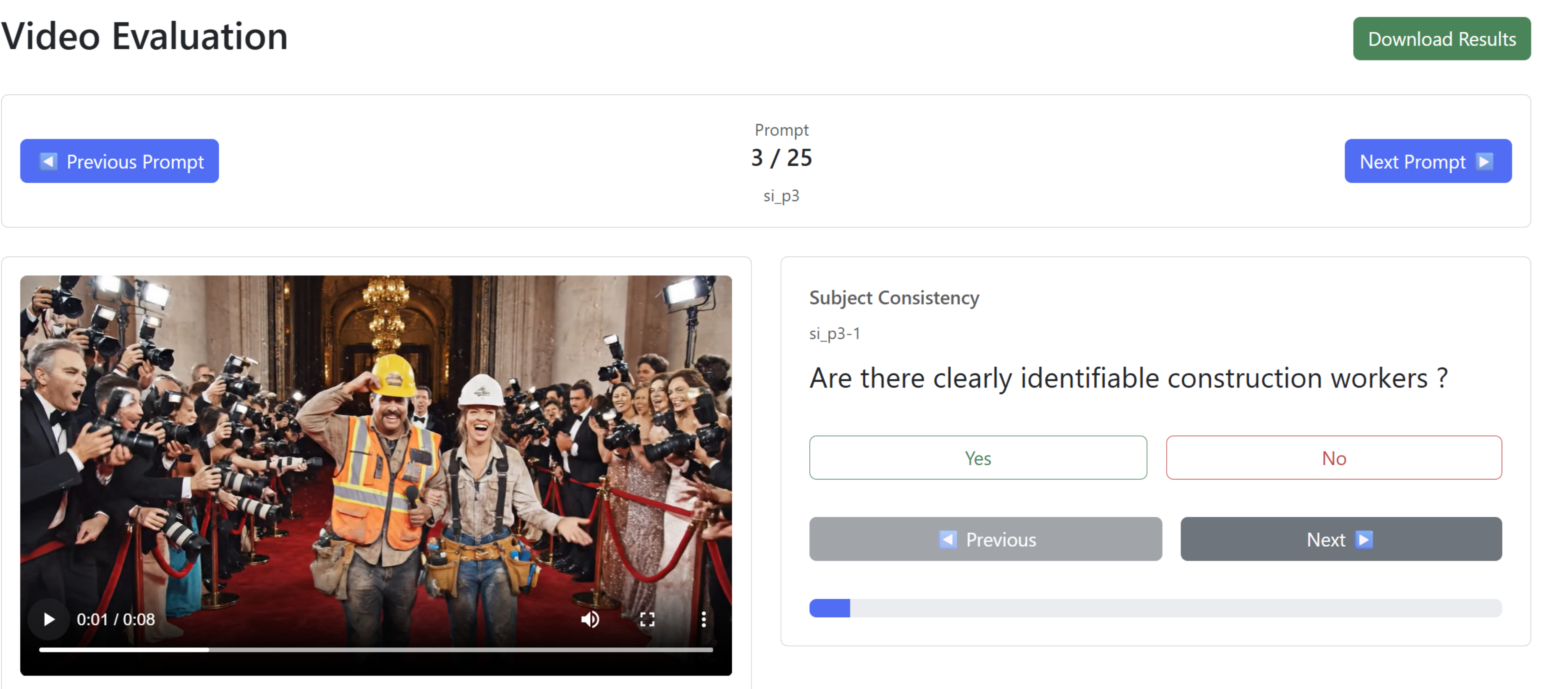}
    \hfill
    \caption{User Annotation tool for Video Evaluation}
    \label{fig:interface}
\end{figure}

%\textbf{Exclusion Rules:} Any question targeting indeterminate attributes—details neither explicitly stated nor logically implied (e.g., the specific color of the sky if not mentioned, or the clothing style of a generic human).Through this human-centric question filtering process, each question yields a consistent binary judgment and enables clear identification of the evaluation dimension in which a failure occurs.

\subsection{Human-Centric Question Answering to Improve Reliability of Ground Truth} \label{human_assesment_protocol}

Current VLM-based evaluators are susceptible to circular reasoning because they share similar training distributions and strong real-world priors with generative models. Consequently, they often judge outputs based on learned notions of plausibility rather than strictly verifying the specific (often implausible) prompt requirements. To overcome this limitation, we implemented a rigorous human-in-the-loop verification protocol. We developed the annotation tool to maximize inspection granularity. Figure \ref{fig:interface} showed its web interface. On this inspection page, there is a playback functionality, and the annotators can play the video any number of times. In fact, annotators were explicitly instructed to watch each video multiple times.

\section{Evaluation and Results} \label{final_analysis}

\subsection{Research Questions}
%\wei{done proofread by Wei, please edit using colored font}

%Our evaluation investigated three research questions: 

{\bf RQ1}: To what extent do T2V models exhibit "Semantic Resistance" against unconventional instructions?  %What are the most difficult implausible scenarios for the models?

Here, we investigate whether models prioritize their training priors (statistical probability) over the explicit user prompt (semantic instruction). Specifically, when a prompt conflicts with world knowledge (e.g., “a barking cat”), does the model produce the logical violation correctly, or does it revert to the statistically probable "mean" (e.g., a meowing cat)? By categorizing prompts into Biological, Physical, Social, and Temporal violations, we aim to isolate specific generation weakness. %blind spots.

%\wei{Localization: fine-grain, root causes}expand to answer localization - reasons

{\bf RQ2}: {Does the subject align with the action in implausible scenarios?}

Here, we aim to quantify the performance gap between generating a correct subject and successfully making actions. A significant drop-off suggests that models excel at retrieving memorized subjects but lack the capabilities and knowledge to bind these to out-of-distribution actions.

%\wei{metric: subject correct, action is wrong}

%\wei{audio video consistency}
{\bf RQ3}: Does the model generate the correct sound for the implausible event (e.g., a barking cat)? Does that sound happen exactly when the action occurs? 

Here, we determine whether models treat audio merely as a loose "background atmosphere" or if they can precisely synchronize the sound with the specific visual trigger (e.g., does the barking sound align with the cat's mouth movement?).

\begin{figure*}[htbp!]
    \centering
\includegraphics[width= \linewidth]{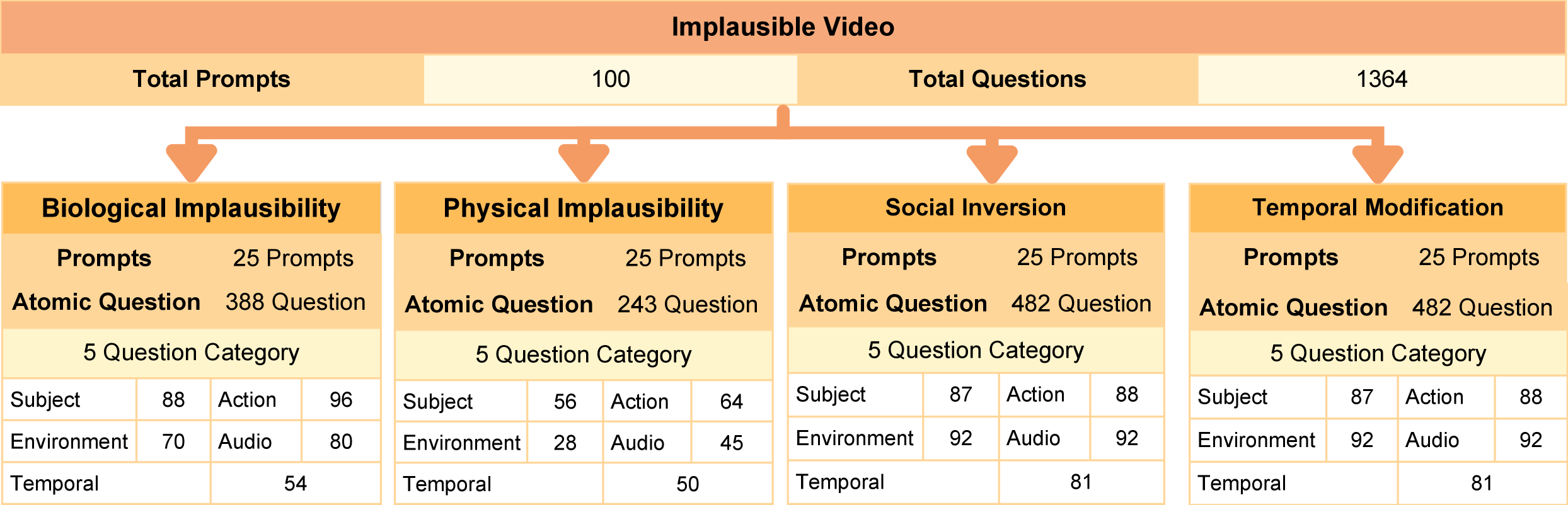}
    \caption{A Reliable and interpretable benchmark generation: 100 prompts for each video generation model and 1364 evaluation questions}
    \label{fig:bm-cons}
\end{figure*}

\subsection{Experimental setup}
\subsubsection{Models}
%\wei{done proofread by wei, using colored font to edit}
We evaluated five state-of-the-art T2V models (Sora 2, Veo 3.1, Runway Gen4.5, PixVerse v5.5, and Qwen3Max) using the BRITE framework: Sora 2 and Veo 3.1 are the best video generation models. Runway Gen 4.5,  despite lacking audio generation, has ranked high in T2V leaderboards. Qwen 3 Max and Pixverse, are among the few models that provided audio generation capabilities. Inclusion of Pixverse, specifically, allows us to analyze how mid-tier models trade off between visual fidelity and audio-visual synchronization compared to the best models like Sora 2 and Veo 3.1.

%\wei{we need to add generation settings here (frames, FPS, resolution, sampling steps, seeds per prompt)}

\subsubsection{Metrics}

We generated 100 prompts for each model, and a total of 1364 evaluation questions covering four categories of implausible scenarios and five dimensions. Figure~\ref{fig:bm-cons} provides the detailed numbers.

\begin{figure}[ht]
\centering
\begin{subfigure}{\linewidth}
\centering
\includegraphics[width=0.90\linewidth]{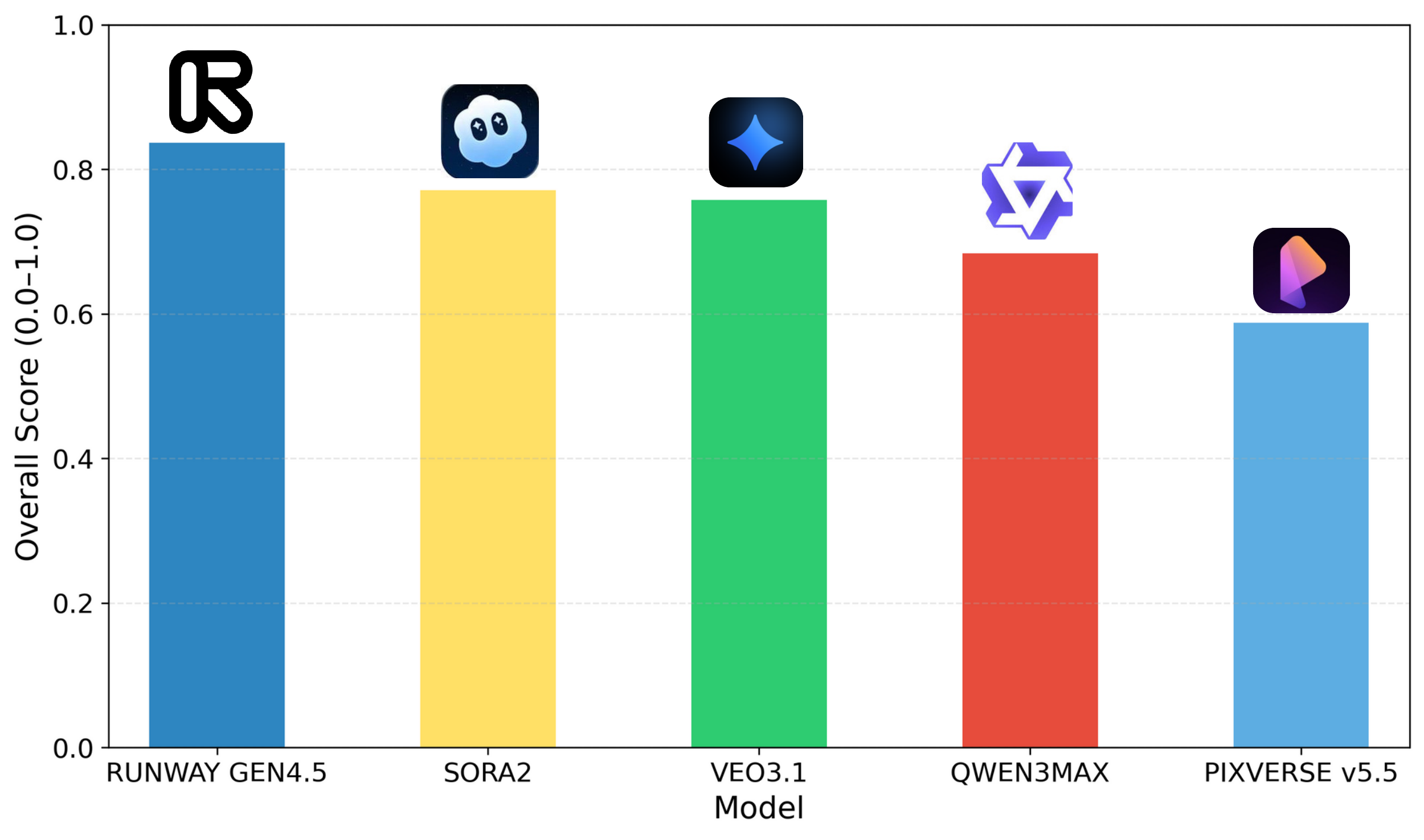}
\caption{Model Performance on Implausible Prompts}
\label{fig:brite_overall}
\end{subfigure}
\hfill
\begin{subfigure}{\linewidth}
\centering
\includegraphics[width=\linewidth]{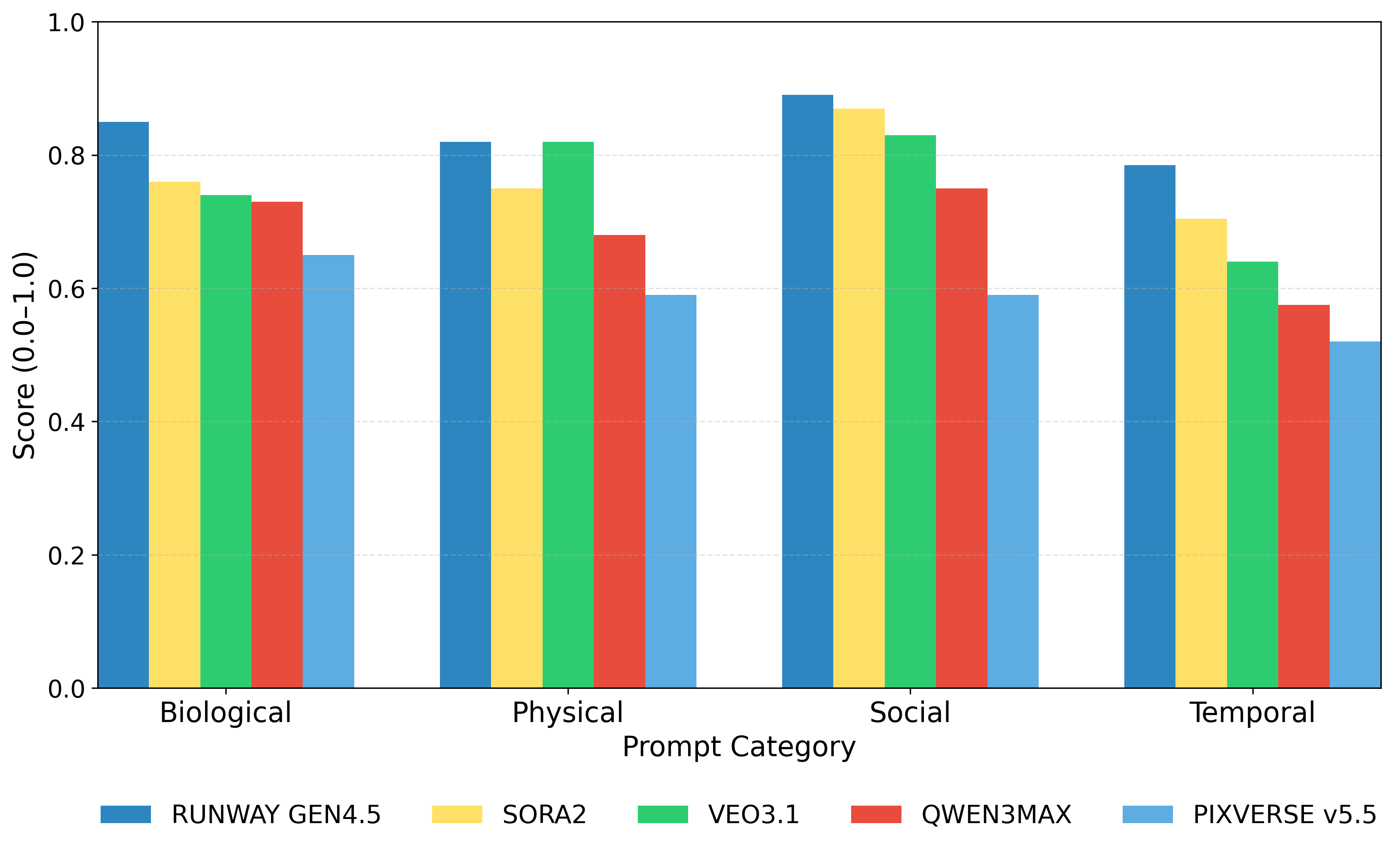}
\caption{Reasoning Performance by Prompt Category}
\label{fig:brite_reasoning}
\end{subfigure}
\caption{BRITE Bench evaluation results across T2V models. The top plot(a) shows overall and prompt adherence performance, while the bottom plot (b) shows reasoning performance across prompt categories.}
\label{fig:brite_combined}
\end{figure}

Compared the ground truth in the prompt with what human annotators have observed in the generated video, we obtain a correctness score for each category. Specifically, {\it subject score} measures if the requested entities (people, objects, animals, plants) are present in the video (e.g., Is it a snail?). {\it Action score} measures if the model correctly executed the actions (e.g., Is the snail moving at an unnaturally fast speed?). {\it Environment score} measures whether the model generated the correct background/scene implied/explicitly requested by the prompt (e.g., savanna for a lion). {\it Audio score} (for multimodal models only) measures whether the model generated the correct type of sound (e.g., Is there a barking sound?). {\it Audio-Visual Sync score} (for multimodal models only) measures if the sounds happen at the exact right time of actions. (e.g., Does the impact sound occur in sync with the visual collision?). 
Table~\ref{tab:implausibility-ablation} reported the number of questions of a particular implausibility category where human answers are consistent with the prompt. In Figure~\ref{fig:brite_combined}, we plotted the data from Table~\ref{tab:implausibility-ablation}. 

We also compared scores for the correlated dimensions in each video.
{\it Subject-Action Gap} measures among all the videos generated by a model, how many videos have a higher subject score than action score (Table~\ref{tab:subject_gt_action}). If this number is high, it indicates the model is good at generating subjects but struggles to create movement. {\it Action-Audio Gap} measures how often a video has the action score higher than the audio score (Table \ref{tab:action_gt_audio}). If this number is low, the model is better at sound than simulating required actions.

\subsubsection{Human Evaluation Protocol}

The evaluation was conducted by two human annotators with extensive prior experience in analysis of generative media. Both annotators possess specialized domain knowledge in identifying generative artifacts (e.g., non-conformity with object permanence etc.) distinct from standard video compression artifacts. To avoid penalizing models for minor, non-semantic artifacts, we adopted the following principles:

\begin{itemize}
    \item \textbf{Failure Threshold:} A "No" (Failure) was recorded only when the video exhibited a clear, definitive violation of the prompt's instructions.
    \item \textbf{Handling Ambiguity:} In "unclear" cases where the violation was ambiguous (e.g., a slightly blurry object), the model was given the benefit of the doubt (marked "Yes"), prioritizing semantic adherence over aesthetics.
    \item \textbf{Tolerances:} Transient artifacts or momentary glitches were disregarded if the core action/subject/audio/synchronization was successfully maintained for the majority of the sequence duration.
\end{itemize}

All videos were annotated independently. In cases of disagreement between the two annotators, questions were flagged and reviewed jointly to reach a final decision, ensuring a unified standard for what constitutes a "failure" across the dataset.

\begin{table*}[t]
\caption{\textbf{Evaluation Results on Four Categories of Implausible Scenarios}}
\centering
\small
\setlength{\tabcolsep}{6pt} % Adjusted for the extra column
\begin{tabular}{lcccccccc}
\toprule
\textbf{Model} 
& \textbf{\shortstack{Video\\Duration}}
& \textbf{Resolution}
& \textbf{\shortstack{Physical\\Implausibility}}
& \textbf{\shortstack{Biological\\Implausibility}}
& \textbf{\shortstack{Social\\Inversion}}
& \textbf{\shortstack{Temporal\\Modification}}
& \textbf{Overall} \\
\midrule

Runway Gen 4.5 & 10s & 720p & 0.82 & {\bf 0.85} & {\bf 0.89} & {\bf 0.79} & {\bf 0.84} \\
Sora 2         & 10s & 720p & 0.75 & 0.76 & 0.87 & 0.70 & 0.77 \\
Veo 3.1        & 8s & 720p & {\bf 0.83} & 0.74 & 0.83 & 0.64 & 0.76 \\
Qwen 3 Max     & 5s & 720p & 0.68 & 0.73 & 0.75 & 0.58 & 0.69 \\
PixVerse 5.5   & 8s & 540p & 0.59 & 0.65 & 0.60 & 0.52 & 0.59 \\
\hline
Average        & & & 0.73 & 0.75 & 0.79 & 0.65 & 0.73 \\

\bottomrule
\end{tabular}
\label{tab:implausibility-ablation}
\end{table*}

%newly added section

\subsection{Results}
%We evaluate five state-of-the-art T2V models (Sora 2, Veo 3.1, Runway Gen4.5, PixVerse v5.5, and Qwen3Max) using the BRITE framework. Our analysis addresses the five core research questions defined at the start of this section, revealing critical failure patterns in how models handle implausibility and multimodal binding.

%For each RQ, we first summarize experimental results, listed under {\bf Data finding}. We then present the insights we draw from the data and examples we see. See {\bf observations}

\subsubsection{RQ1: Model Capabilities on Implausible Scenarios}
%\wei{proofread by wei}
%\textbf{Data Findings:} 
Figure~\ref{fig:brite_overall} showed that Runway Gen4.5 reported best performance in implausible prompts, followed by Sora 2. Among the four implausible categories, temporal modification is the most challenging. The concrete numbers are given in Table~\ref{tab:implausibility-ablation}.  Runway Gen4.5 ranks the top for all the implausiable types except physical implausibility. Runway Gen4.5, Sora 2 and Veo 3.1 all scored above 0.8 for social inversion (such as role reversals).

%reports that Runway Gen4.5 demonstrated the highest overall prompt adherence, leading the benchmark in Biological Implausibility (0.849), Social Inversion (0.888) and Temporal Modification (0.785). The row {\it Average} indicates that Temporal Modification is the most challenging domain for all the models tested. For instance, PixVerse v5.5 and Qwen3Max scored significantly lower in this category (0.520 and 0.575, respectively) compared to others. Interestingly, prompts involving Social Inversion (such as role reversals) yielded the highest success rates across the board, with Sora 2 achieving 0.869 and Runway Gen4.5 reaching 0.888.

%\wei{dimension specific data can be plotted}

%Our evaluation across the four domains of logical violation—Biological Implausibility, Physical Implausibility, Social Inversion, and Temporal Modification—reveals a clear hierarchy of difficulty. Runway Gen4.5 demonstrated the highest overall prompt adherence \textit{(See Table 3)}, leading the benchmark in both Biological Implausibility (0.85) and Social Inversion (0.89). Conversely, Temporal Modification proved to be the most challenging domain for every model tested. For instance, PixVerse v5.5 and Qwen3Max scored significantly lower in this category (0.52 and 0.57, respectively) compared to others. Interestingly, prompts involving Social Inversion (such as role reversals) yielded the highest success rates across the board, with Sora 2 achieving 0.87 and Runway Gen4.5 reaching 0.89.

%\textbf{Observations:} 

These results indicate that models possess uneven "Semantic Resistance." Current architectures struggle significantly more with the physics over time (Temporal Modification) than with the arrangement of social roles. While models can successfully generate a static image of a "toddler instructing a parent" (Social Inversion) because it relies on pixel arrangement, they fail to generate sequences that violate cause-and-effect, such as "shattered glass reassembling." This suggests that while visual composition is robust, the model's understanding of temporal logic remains challenging.

\subsubsection{RQ2: Subject Action Performance Gap}
%\wei{done proofread RQ2, use color font to edit}

%\textbf{Data Findings:}  

In Table~\ref{tab:subject_gt_action} ,  we reveal a measurable gap between generating a subject and animating actions, consistently for all the 5 models. Even for the SOTA models like Sora 2 and Veo 3.1, actions are much more challenging to generate compared to subjects, counting for 63.6\% and 66\% of videos respectively,  especially for the temporal modification scenarios, followed by physical implausibility.

%For example, Sora 2 achieves lower action score in 63.3\% of its videos.Analyzing further,the Subject score exceeds the Action score in over 60\% of generated videos for all models. Furthermore, Table \ref{tab:dim_performance} reveals that Subject scores consistently outperform Action  scores by a wide margin across all models.

%\textbf{Observations:} %This points to a fundamental limitation in current T2V architectures. 
The results imply that T2V models prioritize the high-fidelity rendering of the subject (texture, shape, and identity) but fail to bind these subjects to the requested actions. The models effectively "memorize" what the subject looks like but struggle to simulate how it should move when forced out of their training distribution.

%\wei{example here?}

%\wei{it could be useful to add an example here}

\begin{table}[htbp]
\centering
\footnotesize 
\caption{Subject Action Performance Gap: Percentage of videos where subject score $>$ action score}
\label{tab:subject_gt_action}
\begin{tabularx}{\columnwidth}{lXXXXc}
\toprule
\textbf{Model} & \textbf{Bio.} & \textbf{Phys.} & \textbf{Soc.} & \textbf{Temp.} & \textbf{Total} \\ 
\midrule
Pixverse V5.5  & 12/25 & 22/25 & 9/25  & 22/25 & 65.0\% \\
Qwen3MAX      & 15/24 & 19/25 & 8/25  & 21/25 & 63.6\% \\
Runway 4.5    & 13/25 & 19/25 & 10/25 & 22/25 & 64.0\% \\
Sora2         & 18/25 & 17/23 & 7/25  & 20/25 & 63.3\% \\
Veo3.1        & 19/25 & 18/25 & 8/25  & 21/25 & 66.0\% \\
\bottomrule
\addlinespace[1ex]
\multicolumn{6}{l}{\scriptsize \textit{Prompt Categories: Bio: Biological, Phys: Physical, Soc: Social, Temp: Temporal}} \\
\end{tabularx}
\end{table}

\subsubsection{RQ3: Multimodal Synchronization}

%\textbf{Data Findings:} 

In Table \ref{tab:action_gt_audio}, we show that  Sora 2, Veo3.1 and Qwen3Max generated high-quality, semantically relevant audio. This implies that the models find easier to generate the sound of an implausible scenario than to visually render the dynamics of those actions. However, in the case of PixVerse V5.5, for 60.0\% of its generated videos, the Action score exceeded the Audio score, indicating that its audio generation significantly lags behind its ability to render motion.

\begin{table}[htbp]
\centering
\footnotesize
\caption{Audio and Action Comparison: Percentage of Videos where action score $>$ audio score}
\label{tab:action_gt_audio}
\begin{tabular}{lccccc}
\toprule
\textbf{Model} & \textbf{Bio.} & \textbf{Phys.} & \textbf{Soc.} & \textbf{Temp.} & \textbf{Total} \\ 
\midrule
Pixverse V5.5  & 16/25 & 14/25 & 22/25 & 8/25 & 60.0\% \\
Qwen3MAX      & 6/24  & 5/25  & 16/25 & 8/25 & 35.4\% \\
Sora2         & 3/25  & 7/23  & 7/25  & 6/25 & 23.5\% \\
Veo3.1        & 7/25  & 7/25  & 9/25  & 6/25 & 29.0\% \\
\bottomrule
\addlinespace[1ex]
\multicolumn{6}{l}{\scriptsize \textit{Bio: Biological, Phys: Physical, Soc: Social, Temp: Temporal}} \\
\end{tabular}
\end{table}

%for the majority of implausible scenarios, these advanced models generated appropriate audio  but could not map it to corresponding impossible actions. 

%\textit{(See Table \ref{tab:action_gt_audio})}

\begin{figure}[t]
    \centering
    \includegraphics[width= \columnwidth]{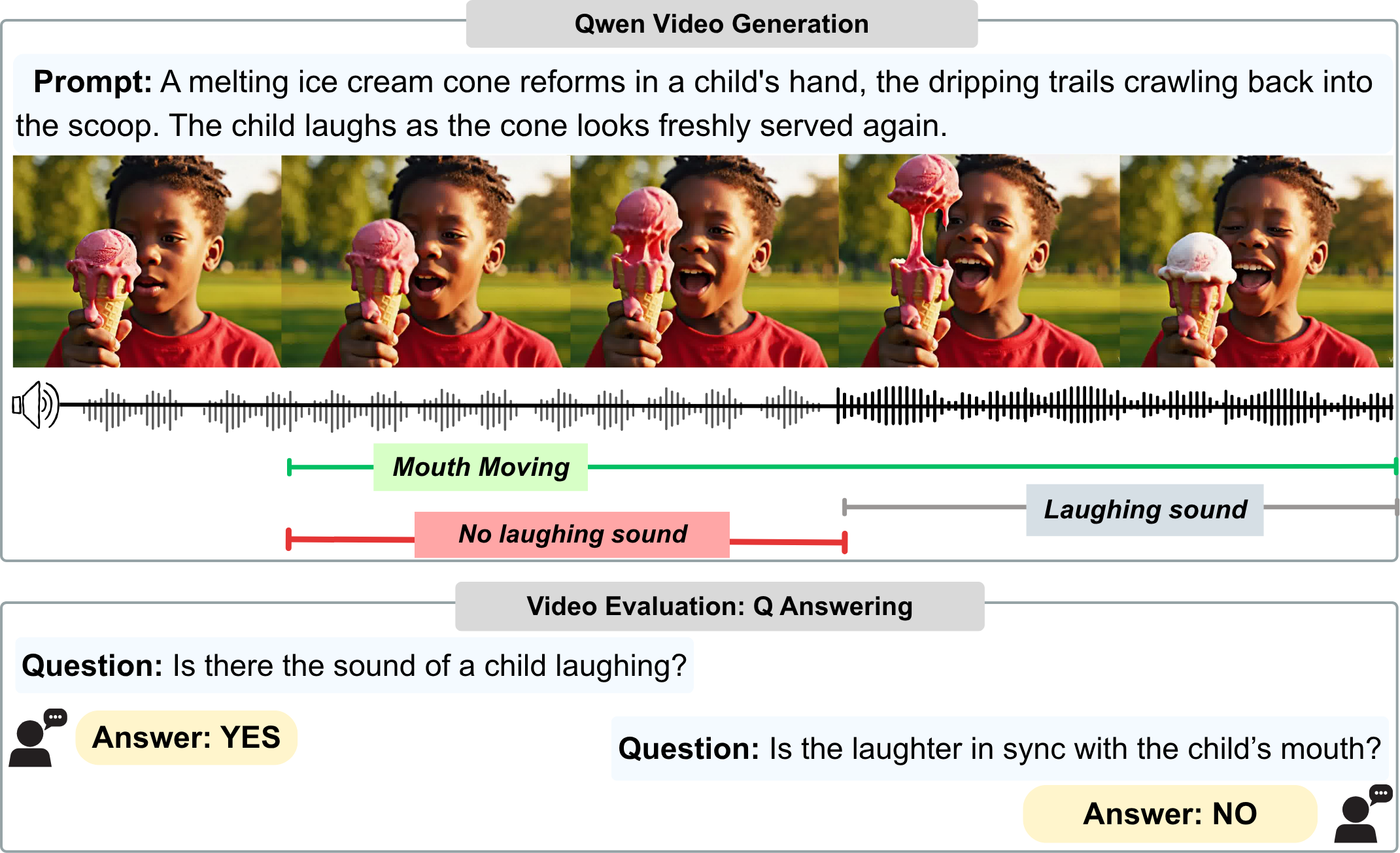}
    \caption{Example of audio–visual mismatch: the laughter sound is not synchronized with the child’s mouth movement.}
    \label{fig:audion_sync}
\end{figure}

Generating the correct sound is only half of the challenge. Table~\ref{tab:dim_performance} showed audio-visual synchronization failed most overall. We observed that current T2V models tend to generate audio primarily as a loose layer on top of the visuals rather than as an action-driven track. In Figure~\ref{fig:audion_sync}, we showed such an example.

\begin{table}[t]
\centering
\caption{\textbf{Evaluation Dimension Comparison across T2V Models}}
\label{tab:dim_performance}

\resizebox{\columnwidth}{!}{
\begin{tabular}{lccccc}
\hline
Model & Subject & Action & Env. & Audio & AV Sync \\
\hline
Runway Gen4.5 & 0.93 & 0.61 & 0.96 &  N/A    & N/A    \\
Sora2         & 0.94 & 0.65 & 0.95 & 0.76 & 0.55 \\
Veo3.1        & 0.92 & 0.58 & 0.97 & 0.69 & 0.63 \\
PixVerse V5.5 & 0.82 & 0.55 & 0.88 & 0.37 & 0.31 \\
Qwen3MAX      & 0.90 & 0.56 & 0.90 & 0.63 & 0.41 \\
\hline
Average       & 0.90 & 0.59 & 0.94 & 0.61 & 0.47 \\
\hline
\end{tabular}
}
\end{table}

% This confirms that while models improve on correctness of audio, the fine-grained audio-video synchronization remains a major area of improvement.

%\wei{proofread this part, it is good}\textbf{Observations:} 

%showed the scores for each dimension for 5 models, calculated as the number of questions human answered yes, which is consistent with the prompt.

%These findings point to two critical conclusions. 

%From the summary data and video we reviewed, we identify that the models find easier to generate the sound of an implausible scenario than to visually render the dynamics of that action. Second, regarding synchronization, we conclude that current T2V models tend to generate audio primarily as a loose layer on top of the visuals rather than as an action-driven track. This confirms that while models improve on correctness of audio fine-grained  audio-video synchronization remains a major area of improvement.

%{While the visual physics degrade under stress, the audio capabilities remain strong. - Can remove - repetitive } 

\section{Related Work}
\textbf{Text-to-Video Generation Models:}
New diffusion and transformer models advance T2V by shifting from silent video to multimodal synthesis. Early frameworks, such as Make-A-Video \cite{singer2022makeavideo} and Imagen-Video \cite{ho2022imagenvideohighdefinition}, optimized pixel-level fidelity but primarily produced silent clips.  VideoPoet \cite{kondratyuk2024videopoet} pioneered the transition toward multimodal synthesis by demonstrating a unified model for joint audio-visual generation. Current frontier models \cite{openai2025sora2, deepmind2025veo3, yang2025qwen25vl, PixVerse2026} now internalize audio as a core component. This shift necessitates evaluation metrics that prioritize audio-visual alignment as a fundamental property rather than a post-processing step. 

\textbf{Taxonomy of T2V Benchmarks:}
Parallel to these architectural gains, benchmarks have transitioned from photorealistic metrics to multidimensional capability suites. EvalCrafter \cite{liu2024evalcrafter} leverages LLM-expanded prompts for everyday scenarios, while VBench/++ \cite{huang2024vbench, huang2024vbenchplus} decomposes quality into hierarchical dimensions like subject consistency and motion stability. Reasoning-focused benchmarks \cite{sun2025t2vcompbench, bansal2025videophy2,chen2025t2vworldbench} assess compositional binding, physics, and world knowledge. However, these datasets often feature plausible scenarios, creating an evaluation "blind spot" where models may appear to follow instructions by retrieving real-world priors. To address this, IPV-Bench
\cite{bai2025impossible} introduces "implausible" scenarios that violate biological or physical rules, revealing a tendency for models to "revert to realism" when prompts contradict training data. 

\textbf{T2V Evaluation methods:}
Methodologies have shifted from low-level statistical distributions to high-level semantic verification. Early metrics like FVD \cite{unterthiner2019accurategenerativemodelsvideo} and IS \cite{tulyakov2017mocogandecomposingmotioncontent} reward "static realism" but fail to capture temporal logic. Although T2V-CompBench \cite{sun2025t2vcompbench} uses detectors for verification of "what is where," newer QA-based frameworks like ETVA \cite{guan2025etva} (for plausible scenarios) and IPV-Bench \cite{bai2025impossible} (for implausible scenarios) rely on Video-LLMs (VLMs) as automated judges. This introduces a circular reasoning loop \cite{bai2025hallucinationmultimodallargelanguage, manduchi2025challengesopportunitiesgenerativeai}: as generators and VLM judges share web-scale training data, the judge often "hallucinates" plausible logic that the prompt required the model to violate. 

Recent stress tests expose the depth of this architectural limitation. VLMs suffer from profound temporal expectation bias, frequently hallucinating physically plausible outcomes rather than grounding their answers in the actual video sequence \cite{t2026stresstestsrevealfragile}. Furthermore, they exhibit directional blindness when physical events are inverted \cite{xue2025seeing, ahn2025happenswhenlearningtemporal}. Most critically, Feng et al. \cite{feng2025breaking} demonstrated that state-of-the-art Video-LLMs exhibit "shuffling invariance"—their performance remains largely unaffected even when video frames are temporally randomized, proving they rely almost entirely on static spatial cues rather than true temporal reasoning. Consequently, in "implausible" scenarios, VLM judges produce severe false positives by defaulting to learned realism. Our work dismantles this circularity bias through a human-centric framework, isolating a model’s ability to override its learned realism priors.

\section{Limitations}
While BRITE provides a rigorous diagnostic framework, we acknowledge the following methodological trade-offs:

Scalability versus Reliability: By prioritizing high-fidelity ground truth over automated evaluation, BRITE relies on human-in-the-loop annotation. Because current Video-LLMs have weak temporal reasoning ability which is required to evaluate implausible scenarios, scoring new models remains a resource-intensive process.

Closed Source Models, Lack of Open-source Models: We evaluated commercial, closed-source models via their consumer-facing chat interfaces using single-seed generations. While evaluating models via their consumer interfaces maximizes ecological validity, it restricts internal interpretability. We can identify the models' spatial and temporal failures, but we cannot trace these errors back to specific internal mechanisms or attention layers.

Dataset Breadth: To achieve deep diagnostic localization, the dataset is constrained to 100 base prompts. We prioritized generating a dense matrix of 1,364 atomic evaluation questions (annotated by two independent experts) over evaluating thousands of random seeds with superficial metrics.

\section{Conclusions and Future Work}
This study presents a reliable, human-in-the-loop framework for testing the limits of state-of-the-art text-to-video models. By evaluating 500 videos across different types of impossible scenarios, we found that current models act more like image compositors than world simulators: they can easily break social norms, but they consistently fail when asked to simulate complex physics or modify the flow of time. Furthermore, while these models often produce high-quality audio that fits the video, they struggle to synchronize it with specific visual actions, revealing a significant gap between generating realistic pixels and understanding the underlying cause-and-effect of the physical world.  These findings suggest that current models focus more on visual fidelity than on causal and temporal correctness. Future research should therefore focus on teaching these systems the rules of the physical world, helping them move beyond videos that merely look realistic on the surface.

\section*{Impact Statement}
The rapid advancement of T2V generation requires up-to-date evaluation. This work advances T2V evaluation by providing a reliable, interpretable benchmark, focusing on implausible scenarios, which only few studies have investigated. This benchmark first introduced audio evaluation, including the correctness of the audio and audio-visual synchronization. The evaluation of the SOTA models located their future areas of improvement, including temporal modification implausible scenarios, and generation of actions, audio, audio-visual synchronization. Our evaluation and framework be extended for rapid future T2V benchmark development.

%whether audio is correct and whether it is aligned with visual.

\bibliography{example_paper}
\bibliographystyle{icml2026}

\section*{Acknowledgments}
We would like to express our deepest gratitude to Dr. Wei Le for her continuous support, guidance, and mentorship throughout the development of this framework. A special thanks is owed to our other contributors who assisted with the evaluation app development, video generation: Ananth Nityandal, Tejas Gosula. We also extend our appreciation to the members of Dr. Le’s Research Group for their insightful discussions and feedback.
%%%%%%%%%%%%%%%%%%%%%%%%%%%%%%%%%%%%%%%%%%%%%%%%%%%%%%%%%%%%%%%%%%%%%%%%%%%%%%%
%%%%%%%%%%%%%%%%%%%%%%%%%%%%%%%%%%%%%%%%%%%%%%%%%%%%%%%%%%%%%%%%%%%%%%%%%%%%%%%
% APPENDIX
%%%%%%%%%%%%%%%%%%%%%%%%%%%%%%%%%%%%%%%%%%%%%%%%%%%%%%%%%%%%%%%%%%%%%%%%%%%%%%%
%%%%%%%%%%%%%%%%%%%%%%%%%%%%%%%%%%%%%%%%%%%%%%%%%%%%%%%%%%%%%%%%%%%%%%%%%%%%%%%
\newpage
\appendix
\onecolumn
% appendix

\section{Video-Generation-Prompt Synthesis Meta-Prompt}\label{app:meta_prompt}
\begin{tcolorbox}[title={LLM-facing Meta-Prompt (lightly edited for clarity)}]
You are an expert prompt designer for text-to-video generation.

\medskip
\textbf{Task.}
Generate video generation prompts that intentionally depict implausible scenarios.
These prompts will be used to evaluate how well video generation models follow
explicit instructions that conflict with common world knowledge.

\medskip
\textbf{4 Implausible Categories.}
\begin{enumerate}
  \item \textbf{Physical Implausibility:}
  Violates physical laws or material properties
  
  (e.g., a feather falling faster than a bowling ball).
  \item \textbf{Biological Implausibility:}
  Organisms act against biological traits or instincts
  
  (e.g., a barking cat; plant roots growing upward into the sky).
  \item \textbf{Temporal Modification:}
  Alters the normal flow of time (reversal, slow motion, acceleration)
  
  (e.g., shattered glass reassembling itself).
  \item \textbf{Social Inversion:}
  Reverses established social roles or norms
  
  (e.g., a wedding where everyone wears gym clothes).
\end{enumerate}

\textbf{Constraints.}
\begin{itemize}
  \item Generate 50 prompts per category.
  \item Each prompt should be no longer than two sentences.
\end{itemize}

\textbf{Example.}
``A cat barking loudly while sitting on a kitchen counter.''

\textbf{Output.}
Provide the prompts grouped by category.
\end{tcolorbox}

\section{Question-Generation Meta-Prompt}\label{app:question_generation}

\begin{tcolorbox}[title={LLM-facing Meta-Prompt (lightly edited for clarity)}]

You are an expert evaluator for text-to-video generation.

\medskip
\textbf{Task.}
Given a set of video generation prompts belonging to a single implausible category
(Physical Implausibility, Biological Implausibility, Temporal Modification, or
Social Inversion), generate evaluation questions for each prompt to assess how
well a generated video follows the prompt.

\medskip
\textbf{5 Evaluation Dimensions.}
\begin{enumerate}
    \item \textbf{Subject Consistency:}
    Are the subjects consistent with the prompt?

    \item \textbf{Action Consistency:}
    Are the actions consistent with the prompt?

    \item \textbf{Environment Consistency:}
    Is the environment consistent with the prompt?

    \item \textbf{Audio Consistency:}
    Is the audio consistent with the prompt?

    \item \textbf{Audio-Visual Synchronization:}
    Are audio events synchronized with visual events?
\end{enumerate}

\medskip
\textbf{Instructions.}
\begin{itemize}
    \item Generate 3--5 questions per evaluation dimension.
\end{itemize}

\medskip
\textbf{Output Format.}
Return the questions in a structured JSON format.

\medskip
The JSON output serves as an evaluation questionnaire template, with answer
fields initialized as \texttt{null} and filled later by human annotators or
evaluation models.

\medskip
\textbf{Processing.}
Each prompt is processed independently (prompt-by-prompt).

\end{tcolorbox}

%%%%

\end{document}